\documentclass[12pt]{article}

\newlength{\extraspace}
\newlength{\extraspaces}
\newcommand{\newsection}[1]{
\vspace{7mm}
\pagebreak[3]
\addtocounter{section}{1}
\setcounter{equation}{0}
\setcounter{subsection}{0}
\setcounter{footnote}{0}
\begin{center}
{\large {\bf \thesection. #1}}
\end{center}
\nopagebreak
\medskip
\nopagebreak
\hspace{3mm}}

\newcommand{\be}{\begin{equation}
\addtolength{\abovedisplayskip}{\extraspaces}
\addtolength{\belowdisplayskip}{\extraspaces}
\addtolength{\abovedisplayshortskip}{\extraspace}
\addtolength{\belowdisplayshortskip}{\extraspace}}
\newcommand{\ee}{\end{equation}}

\newcommand{\ba}{\begin{eqnarray}
\addtolength{\abovedisplayskip}{\extraspaces}
\addtolength{\belowdisplayskip}{\extraspaces}
\addtolength{\abovedisplayshortskip}{\extraspace}
\addtolength{\belowdisplayshortskip}{\extraspace}}
\newcommand{\ea}{\end{eqnarray}}

\def\nn{\nonumber \\}
\def\txt{\textstyle}

\newcommand{\VEV}[1]{\left\langle {#1} \right\rangle}

\begin{document}
\addtolength{\baselineskip}{.7mm}
\thispagestyle{empty}

\begin{flushright}
RIKEN--TH--28 \\
STUPP--04--175 \\
June, 2004
\end{flushright}
\vspace{.2cm}

\begin{center}

{\large{\bf{
Minimal String Theory is Logarithmic
}}}
\\[12mm]

{\sc Yukitaka Ishimoto}
$^{\rm a}$\,\footnote{\ E-mail: y.ishimoto@riken.jp}
\quad and \quad 
{\sc Shun-ichi Yamaguchi}
$^{\rm b,c}$\,\footnote{\ E-mail: syama@post.saitama-u.ac.jp}
\\[10mm]

$^{\rm a}$
{\it 
Theoretical Physics Laboratory, RIKEN, \\
Wako 351-0198, Japan
}
\\[5mm]

$^{\rm b}$
{\it 
Department of Physics, Faculty of Science, Saitama University, \\
Saitama 338-8570, Japan
}
\\[2mm]

$^{\rm c}$
{\it 
Department of Physics, School of Science, Kitasato University, \\
Sagamihara 228-8555, Japan
}
\\[15mm]

{\bf Abstract}
\\[7mm]
{\parbox{13cm}{
We study the simplest examples of minimal string theory whose worldsheet description is the unitary $(p,q)$ minimal model coupled to two-dimensional gravity (\nolinebreak{Liouville} field theory). In the Liouville sector, we show that four-point correlation functions of `tachyons' exhibit logarithmic singularities, and that the theory turns out to be logarithmic. The relation with Zamolodchikov's logarithmic degenerate fields is also discussed. Our result holds for generic values of $(p,q)$.
}}
\\[10mm]

\end{center}

\begin{footnotesize}



\end{footnotesize}


\newpage
\setcounter{section}{0}
\setcounter{equation}{0}


\newsection{
Introduction
}
Minimal string theories are interesting string laboratories whose target space is two dimensions (see \cite{Ginsparg:is,Se04} and the references therein).
The total central charge is 26 for bosonic cases, and 
their worldsheet description can be realised by two-dimensional gravity (Liouville field theory) coupled to $(p,q)$ minimal conformal field theories \cite{BPZ,DDK}. 

In the simplest examples of such theories, the Liouville sector and the $(p,q)$-matter sector almost decouple from each other, except that the on-shell condition provides a bridge between the two sectors. `Tachyon' is such a field that satisfies the condition and is a tensor product of a Liouville primary and a $(p,q)$ primary. In \cite{SY02}, one of the authors has shown that such a theory with $(p,q)=(4,3)$ has a peculiar Liouville four-point function which exhibits logarithmic singularity. In this paper, we find that quite many four-point functions of tachyons have logarithmic singularities for generic $(p,q)$, and that minimal string theory is logarithmic.
This fact implies that all the $(p,q)$ minimal string theories are logarithmic theory with logarithmic fields.

Logarithmic conformal field theory (LCFT) is a class of conformal field theories (CFTs) which was first found in a $c=-2$ theory in \cite{Gurarie} from four-point functions with logarithms \cite{Knizhnik:xp}. The main feature of this theory is a reducible but indecomposable representation, or Jordan cell structure, and its corresponding logarithmic fields [8--14].
Soon after \cite{Gurarie}, another example of LCFT was shown to be a gravitationally dressed CFT, where fermionic four-point functions had logarithmic terms in the vicinity of $\xi=1$ \cite{Bilal:1994nx}. Here, $\xi$ is an harmonic ratio. It was also suggested that a puncture-type operator appearing in Liouville field theory may play a role in LCFT as a pre-logarithmic field \cite{Kogan:1997fd}. Liouville field theory has been involved in this game. 

Zamolodchikov has recently shown an operator-valued relation of logarithmic degenerate fields with a particular type of primaries \cite{Za03}. Such primaries naturally emerge in two-dimensional gravity coupled to minimal matter.
It has been suggested in \cite{logWZW} that Liouville correlation functions serve four-point functions of $SL(2,{\mathbf R})$ WZNW models.
On the other hand, one of the authors has calculated chiral four-point functions and conditions for logarithms in the Coulomb gas picture with a boundary in \cite{Ishimoto:2003nb}. 

In the rest of this paper, 
we calculate Liouville four-point correlation functions of the gravitational sector of tachyons in two-dimensional gravity coupled to $(p,q)$ minimal conformal field theories without applying the differentiation method \cite{SY02,SY00}. 
We also aim to show the solutions in a generic form 
to enable study of the non-trivial Liouville dynamics and a more explicit relation between two-dimensional gravity and LCFT.
It should be noted that, unlike in \cite{Bilal:1994nx}, the conformal gauge is used here and our results contain logarithms at both $\xi=0$ and $\xi=1$.


\newsection{
Action, tachyons, and Liouville correlation functions
}
We start with two-dimensional gravity coupled to 
the $(p,q)$ minimal model. 
In the conformal gauge, the gravitational sector is described by 
Liouville field theory on a sphere with the action \cite{DDK}:
\be
S_{\rm L}[\hat g, \phi] = 
\frac{1}{8\pi} \int d^2 z \sqrt{\hat g} \left(
\hat g^{\alpha \beta} \partial_\alpha \phi \partial_\beta \phi 
- Q \hat R \phi + 4 \mu \,e^{\alpha \phi} \right), 
\label{action}
\ee
where $\hat R$ is the two-dimensional scalar curvature 
with the fixed reference metric $\hat g_{\alpha \beta}$ and 
$\mu$ is the renormalised cosmological constant. 
The parameter $Q$ is defined by 
$Q = - \alpha - \frac{2}{\alpha}$ with $\alpha = -\sqrt\frac{2q}{p}$. 
In this case, Liouville field theory becomes CFT 
with the central charge $c_{\rm L} = 1 + 3 Q^2$ \cite{CT}. 

The `tachyon' operator is defined by \cite{DDK,DFKu}:
\be
O_{r,t} = \int d^2 z \sqrt{\hat g} \, O_{r,t} (z, \bar z), 
\label{physop}
\ee
with the on-shell condition that the total conformal weight of $O_{r,t}(z, \bar z)$ is one.
Here, $O_{r,t} (z, \bar z)$ is called the gravitationally dressed operator: 
\be
O_{r,t}(z, \bar z) 
= e^{\beta_{r,t} \phi(z, \bar z)} \Phi_{r,t}(z, \bar z)
\label{dressed}
\ee
and $\Phi_{r,t}(z, \bar z)$ is a Kac primary field 
of the $(p,q)$ matter \cite{BPZ}. 
Therefore, $(r,t)$ is restricted in a rectangular region, 
and the value of $\beta_{r,t}$ is fixed by the on-shell condition \cite{SePo}: 
\be
\beta_{r,t} = (1-r) \frac{1}{\alpha} + (1+t) \frac{\alpha}{2}, 
\label{betars}
\ee
since 
the conformal weight of the Liouville primary operator, $e^{\beta \phi(z, \bar z)}$, is $h_\beta =\nolinebreak -\frac{1}{2} \beta^2 - \nolinebreak\frac{1}{2} \beta Q$. 
%
Note that, in \cite{Za03}, 
Liouville primaries with the following 
condition appear in our notation in an operator-valued relation of logarithmic 
degenerate fields: 
\be
\beta_{m,n}
= (1-m) \frac{1}{\alpha} +(1+n) \frac{\alpha}{2}. 
\ee
This condition coincides with the on-shell condition, 
{\it i.e.}\ eq.\ (\ref{betars}). 

As the matter sector is well studied and known, we will neglect it 
and consider only the Liouville part of four-point correlation functions 
of operators (\ref{physop}) \cite{DFKu,LVEV}. After integrating out 
the Liouville zero mode $\phi_0$ $(\phi = \phi_0 + \tilde \phi)$, 
the Liouville correlation function on the complex plane becomes: 
\be
\VEV{\prod_{i=1}^4 e^{\beta_i \phi(z_i, \bar z_i)}}
= \left(\frac{\mu}{2 \pi} \right)^{\! s} 
\frac{\Gamma (-s)}{-\alpha} \, \tilde G_{\rm L}^{(s)}, 
\label{GL}
\ee
where the function $\tilde G_{\rm L}^{(s)}$ is the non-zero mode expectation value:
\be
\tilde G_{\rm L}^{(s)} 
= \VEV{\prod_{i=1}^4 e^{\beta_i \tilde \phi(z_i, \bar z_i)} 
\left(\int d^2 u \,e^{\alpha \tilde\phi(u, \bar u)}\right)^{\! s}}, 
\label{afterzero}
\ee
with the free field action of $\tilde \phi$, 
and the parameter $s$ is given by:
\be
s = -\frac{1}{\alpha} \left( Q + \sum_{i=1}^4 \beta_{i} \right). 
\label{s}
\ee
$\beta_i$ denotes $\beta_{r_i,t_i}$. 
The calculation of the function (\ref{GL}) is analogous to that of 
the Coulomb gas picture \cite{DF}. 
In case $s$ be non-negative integer, one has to 
interpret a singular factor in eq.\ (\ref{GL}) 
as $\left(\frac{\mu}{2\pi} \right)^{s} \Gamma (-s)$ 
be 
$\left(\frac{\mu}{2 \pi} \right)^{s} \frac{(-1)^{s+1}}{\Gamma(s+1)} 
\ln \mu$ \cite{DFKu}. 

When $s$ is a non-negative integer, we can 
evaluate eq.\ (\ref{afterzero}) as shown in \cite{DF,SY02}:
\ba
\tilde G_{\rm L}^{(s)}
& \!\!\! = & \!\!\! \prod_{1 \leq i < j \leq 4} 
|z_i - z_j|^{-2(h_i +h_j) + \frac{2}{3} h} \, 
|\xi|^{2(h_1 + h_2) - \frac{2}{3} h -2 \beta_1 \beta_2}
\nonumber \\[.5mm]
& \!\!\! & \!\!\! \times \, |1- \xi|^{2(h_2 + h_3) - \frac{2}{3} h 
-2 \beta_2 \beta_3} \, 
I^{(s)}(-\alpha\beta_1, -\alpha\beta_3, -\alpha\beta_2; 
- {\textstyle\frac{1}{2} \alpha^2}; \xi, \bar\xi), 
\ea
where $h = \sum\limits_{i=1}^4 h_i$, 
$h_i = h_{\beta_i}$, 
$\xi =\frac{(z_1 -z_2)(z_3 -z_4)}{(z_1 -z_3)(z_2 -z_4)}$ and 
\vspace{-2pt}
\be
I^{(s)} (a, b, c; \rho; \xi, \bar\xi) 
= \int \prod_{i=1}^s d^2 u_i \prod_{i=1}^s 
\left[ |u_i|^{2a} |1-u_i|^{2b} |u_i-\xi|^{2c} \right] 
\prod_{1 \leq i < j \leq s} |u_i - u_j|^{4\rho}. 
\label{Is}
\ee
When $s=0$, the integral (\ref{Is}) vanishes 
and $\tilde G_{\rm L}^{(0)}$ turns out to be a simple product of powers. 
In the following sections, 
we will consider the next-to-trivial case, that is, $s=1$, 
where the integral (\ref{Is}) becomes: 
\be
I^{(1)}(-\alpha\beta_1, -\alpha\beta_3, -\alpha\beta_2; 0; 
\xi, \bar\xi)
\,=\, \int d^2 u \,|u|^{-2 \alpha\beta_1} |1-u|^{-2 \alpha\beta_3} 
|u-\xi|^{-2 \alpha\beta_2}. 
\label{s1int}
\ee
As has been pointed out in \cite{SY02}, eq.\ (\ref{s1int}) may have logarithmic terms.


\newsection{
Four fields for $s=1$
}
When $p$ and $q$ are coprime, $s$ is generally a fractional number. However, for appropriate $p$ and $q$, there exist such a combination of four fields that gives $s=1$. 
Namely, 
we state that, given that $p,q,r,t \in \mathbf{Z}$, $p$ and $q$ are coprime, and 
\be
p \geq 4,\ q \geq 3, 
\label{pq}
\ee
then there exist such combinations of fields that give $s=1$:
\be
\VEV{O_{r,t}\, O_{q-(r-1),p-(t+2)}\, O_{r,t}\, O_{q-(r-1),p-(t+2)}}  
\quad  {\rm for~~} 2 \leq r \leq q-1,\ 1 \leq t \leq p-3. 
\label{combi}
\ee

Kac primaries of the matter part reside only in the conformal grid ${\mathbf G}_{p,q}$:
\be
{\mathbf G}_{p,q} = \left\{(r,t) \in {\mathbf Z}^2 
\,\vert\, 1 \leq r \leq q-1,\ 1 \leq t \leq p-1 \right\}, 
\ee
so do the operators $\{ O_{r,t} \}$. The existence of the combinations in (\ref{combi}) means this requirement which is realised by $2\leq r\leq q-1$, $1\leq t \leq p-3$ under the condition (\ref{pq}).
Thus, one can easily see that the statement is true. Note that $p\geq 2$, $q\geq 2$ for $\mathbf{G}_{p,q} \neq \emptyset$, and that the number of such combinations is $(p-3) (q-2)$.

One can also show that, if we assume that $p, q$ are coprime and $s=1$ for the $\VEV{ABAB}$ type of correlation functions, the only possible combinations are those shown in (\ref{combi}) provided that $p$, $q$ satisfy the inequalities (\ref{pq}).

It follows from $\VEV{ABAB}$ that $\beta_1=\beta_3$ and $\beta_2=\beta_4$.
With $s=1$ and $\beta_i = (1-r_i)\frac{1}{\alpha} + (1+t_i)\frac{\alpha}{2}$, the relation (\ref{s}) reduces to 
$\sum_{i=1}^2 r_i -1 = \frac{q}{p} \left( \sum_{i=1}^2 t_i +2 \right)$. 
Since $p, q$ are coprime, the sum in the parentheses on the {\it r.h.s.}\ should be a multiple of $p$. Therefore,\footnote{\
For non-coprime $p$, $q$ cases, $(p,q)$ are replaced by the greatest coprime divisors of $p$ and $q$.
}
$$
   \sum_{i=1}^2 r_i = n q +1 , \quad
   \sum_{i=1}^2 t_i = n p -2 , \quad {\rm for~~} n\in\mathbf{Z}. 
$$
Since $(r_i,t_i)\in \mathbf{G}_{p,q}$, $2\leq\sum r_i\leq 2q-2$ and $2\leq\sum t_i \leq 2p-2$. Therefore, $n=1$ is the only choice. Hence,
$$
   r_2 = q - (r_1-1) , \quad
   t_2 = p - (t_1+2) . 
$$
Since $(r_i,t_i)\in \mathbf{G}_{p,q}$, this leads to the combinations in (\ref{combi}) and the inequalities (\ref{pq}).


\newsection{
Integral expressions and explicit calculations
}
Let us consider the Liouville part of the following correlation functions:
\begin{eqnarray}
\VEV{O_{r,t}(z_1, \bar z_1)\,O_{q-(r-1), p-(t+2)}(z_2, \bar z_2)\,
O_{r,t}(z_3, \bar z_3)\,O_{q-(r-1), p-(t+2)}(z_4, \bar z_4)}, 
\label{ex0}
\end{eqnarray}
for $2\leq r\leq q-1,\ 1\leq t \leq p-3$ with the inequalities (\ref{pq}).  
Note that (\ref{pq}) is not odd, since it is only a unitarity bound for the minimal models.

Firstly, we should calculate the following integral for $s=1$:
\be
I^{(1)} (a,b,c;0;\xi,\bar \xi) 
= \int d^2 u \,|u|^{2a} |1-u|^{2b} |u-\xi|^{2c}.
\label{int}
\ee
In ordinary CFT, this can be reduced to the following form by Dotsenko's formula \cite{D88}:
\be
I^{(1)} (a,b,c;0;\xi,\bar \xi) = G_{1}\,|F_1(\xi)|^2 + G_{2}\,|F_2(\xi)|^2,
\label{dotsenko}
\ee
where $G_i$s are $\xi$-independent functions of $a, b$ and $c$, and $F_i(\xi)$s consist of two independent hypergeometric functions.
However in some cases, or LCFT cases in particular, this form may be indefinite, having vanishing denominators in $G_i$s. This is to be treated carefully, because if one wants to pursue this form, it is necessary to bring some nontrivial techniques like the differentiation method in \cite{SY02,SY00}.

There is another way of expressing the integral in terms of hypergeometric functions. This avoids such indefinite forms and make the formula directly applicable to general cases, including LCFT cases. The procedure simply involves performing the same analytic continuation as that in \cite{D88}, then expressing two out of four integrals in another domain $|\xi-1|<1$ as follows:
\be
I^{(1)} (a,b,c;0;\xi,\bar \xi) 
= - \sin (\pi a) \,I_2(\xi) \,I_3(\bar \xi) 
- \sin (\pi b) \,I_4(\xi) \,I_1(\bar \xi), 
\label{new I}
\ee
where
\ba
I_1(\xi) & \!\!\! \equiv & \!\!\! 
\int_1^\infty du \, u^a (u-1)^b (u-\xi)^c
\nonumber \\[.5mm]
& \!\!\! = & \!\!\! 
\frac{\Gamma(-1-a-b-c)\,\Gamma(1+b)}{\Gamma(-a-c)} \,
{}_2 F_1 (-c, -1-a-b-c; -a-c; \xi), 
\nonumber \\[.5mm]
I_2(\xi) & \!\!\! \equiv & \!\!\! 
\int_0^\xi du \, u^a (1-u)^b (\xi-u)^c
\nonumber \\[.5mm]
& \!\!\! = & \!\!\! 
\frac{\Gamma(1+a)\,\Gamma(1+c)}{\Gamma(2+a+c)} \,
\xi^{1+a+c} \,
{}_2 F_1 (-b, 1+a; 2+a+c; \xi), 
\nonumber \\[.5mm]
I_3(\xi) & \!\!\! \equiv & \!\!\! 
\int_{-\infty}^0 du \, (-u)^a (1-u)^b (\xi-u)^c
=
\int_1^\infty du \, (u)^a (u-1)^b (u-(1-\xi))^c
\nonumber \\[.5mm]
& \!\!\! = & \!\!\! 
\frac{\Gamma(-1-a-b-c)\,\Gamma(1+a)}{\Gamma(-b-c)} \,
{}_2 F_1 (-c, -1-a-b-c; -b-c; 1-\xi), 
\nonumber \\[.5mm]
I_4(\xi) & \!\!\! \equiv & \!\!\! 
\int_\xi^1 du \, u^a (1-u)^b (u-\xi)^c
=
\int_0^{1-\xi} du \, (u)^b (1-u)^a (1-\xi-u)^c
\nonumber \\[.5mm]
& \!\!\! = & \!\!\! 
\frac{\Gamma(1+b)\,\Gamma(1+c)}{\Gamma(2+b+c)} \,
(1-\xi)^{1+b+c} \,
{}_2 F_1 (-a, 1+b; 2+b+c; 1-\xi). 
\label{I1234}
\ea
Substituting the above into eq.\ (\ref{new I}), we obtain:
\ba
\hspace{-8mm}
I^{(1)} (a,b,c;0;\xi,\bar \xi) 
& \!\!\! = & \!\!\! 
- \,\Gamma(-1-a-b-c) \,\Gamma(1+c)
\nonumber \\[.5mm]
& \!\!\! & \!\!\! 
\hspace{-32mm}
\times \,\Bigl[\,\sin(\pi a) \,U_{23} \,\xi^{1+a+c} \,
\nonumber \\[.5mm]
& \!\!\! & \!\!\! 
\hspace{-32mm} \quad \ \quad 
\times \,{}_2 F_1 (-b, 1+a; 2+a+c; \xi) 
\,{}_2 F_1 (-c, -1-a-b-c; -b-c; 1- \bar \xi)
\nonumber \\[.5mm]
& \!\!\! & \!\!\! 
\hspace{-32mm}
\quad \ 
+ \, \sin(\pi b) \,U_{41} \,(1-\xi)^{1+b+c} \,
\nonumber \\[.5mm]
& \!\!\! & \!\!\! 
\hspace{-32mm} \quad \ \quad 
\times \,{}_2 F_1 (-a, 1+b; 2+b+c; 1-\xi)
\,{}_2 F_1 (-c, -1-a-b-c; -a-c; \bar \xi) \Bigr], 
\label{intrep}
\ea
where $U_{23} 
= \frac{[\Gamma(1+a)]^2}{\Gamma(2+a+c) \,\Gamma(-b-c)}$ and 
$U_{41} 
= \frac{[\Gamma(1+b)]^2}{\Gamma(2+b+c) \,\Gamma(-a-c)}$. 

For the integrals in (\ref{I1234}) to be of the hypergeometric functions, the integrands should not have poles at 
$u=0,1,\xi,\infty$, {\it i.e.}\ $a,b,c \not\in {\mathbf Z_-}$.
Therefore, the above formula (\ref{intrep}) is valid for such values of $a,b$ and $c$.
When $(1+a+c) \not\in {\mathbf Z}$ and $(1+b+c) \not\in {\mathbf Z}$, one can easily see that this is equivalent to Dotsenko's formula of the type (\ref{dotsenko}).
It should also be mentioned here that, precisely speaking, the analytic continuation is well-defined and exact when $(a+b+c)<-1$ and $\xi$ is real.
However, the integral (\ref{int}) and {\it r.h.s.}\ of the formula (\ref{intrep}) do not appear to be ill-defined nor singular when the value of $\xi$ deviates a little from the real axis. So we can simply assume that $\xi$ can be analytically continued to the whole complex plane, or at least, eq.\ (\ref{intrep}) can be regarded as a regularised expression of the integral (\ref{int}).

By using the foregoing formula, one can explicitly calculate the Liouville part of the functions (\ref{ex0}). 
Since the formula (\ref{intrep}) and 
\ba
- \alpha \beta_1 &\!\!\! \equiv &\!\!\! A = -1 + r - (1+t) \frac{q}{p}, 
\nn
- \alpha \beta_2 &\!\!\! = &\!\!\! - r + (1+t) \frac{q}{p} = -1-A, 
\ea
the two-dimensional integral (\ref{s1int}), or $I^{(1)}(-\alpha \beta_1, - \alpha \beta_3, - \alpha \beta_2, 0 ;\xi,\bar \xi)$, amounts to:
\ba
\hspace{-27pt} &\!\!\! &\!\!\! \hspace{-20pt}
I^{(1)}(A,A,-1-A; 0; \xi,\bar \xi)
\nonumber \\[.5mm]
\hspace{-27pt} &\!\!\! =&\!\!\! 
\frac{(-1)^{1+r} \pi^2}{\sin \left(\pi (1+t) \frac{q}{p} \right)}
\left\{ {}_2 F_1 (-A,1+A;1;\xi) \,{}_2 F_1 (-A,1+A;1; 1-\bar \xi) + (c. \,c.)\right\}.
\label{result0}
\ea
Therefore, we obtain $\tilde G_{\rm L}^{(1)}$ of eq.\ (\ref{GL}) as 
\ba
\tilde G_{\rm L}^{(1)}
= |z_1-z_3|^{-4 h_1} |z_2-z_4|^{-4 h_2} |\xi(1-\xi)|^{-2 \beta_1 \beta_2} I^{(1)}(A,A,-1-A; 0; \xi,\bar \xi).
\ea
Here we use the gamma function identity, 
$\Gamma(-x) \, \Gamma(1+x) = - \pi/\sin(\pi x)$. 
The integral is $\xi \leftrightarrow \bar \xi$ symmetric and therefore real, as it should be. In other words, it is single-valued and monodromy invariant.
The conditions ($a, b, c \not\in \mathbf{Z}_-$) for the formula (\ref{intrep}) 
are guaranteed by non-integer values of $A$, since $(1+t)\frac{q}{p} \not\in {\mathbf Z}$ with $t\leq p-3$ and coprime $(p,q)$.
Note that we do not strictly apply the condition $(a+b+c)<-1$ of (\ref{intrep}), although, when $q<p<2q$, the condition allows all the possible pairings of $O_{r,t} \, O_{q-(r-1), p-(t+2)}$.
In particular, $(r,t)=(2,1)$ is allowed when $p<2q$, and $q<p<2q$ with eq.\ (\ref{pq}) includes all the non-trivial unitary minimal $(q+1, q\geq 3)$ models. 

Hence, we find that the Liouville parts of the correlation functions (\ref{ex0}) have logarithmic terms in (\ref{result0}), since 
\pagebreak
%
\ba
&\!\!\! &\!\!\! \hspace{-25pt}
{}_2 F_1 (-A,1+A;1;\xi) 
\nopagebreak\nonumber \\[.5mm]\nopagebreak
\hspace{-25pt} &\!\!\! = &\!\!\! 
\frac{\sin(\pi A)}{\pi} \Biggl[\,\ln (1-\xi) \, {}_2 F_1 (-A,1+A;1; 1-\xi)
\nonumber \\[.5mm]
\hspace{-25pt} &\!\!\! {} &\!\!\! 
-\sum_{n=0}^{\infty} \frac{(-A)_n (1+A)_n}{(n!)^2} \,(1-\xi)^n
\Bigl\{ 2 \psi(n+1) - \psi(-A+n) - \psi(1+A+n) \Bigr\}\Biggr], 
\nonumber \\
\label{log}
\ea
where $(a)_n$ is the Pochhammer symbol and $\psi(x) = \frac{\partial}{\partial x} \ln(\Gamma(x))$. 
This fact further demonstrates that the theory should contain logarithmic operators whose two-point functions yield logarithmic terms. In a chiral theory, such operators can be seen in the operator algebra as follows \cite{Gurarie}:
\be
O_1(z) \, O_2(0) \,\sim \,z^{-h_1-h_2+h_C} \left( C(0) \ln(z) + D(0) \right),
\ee
with the following two-point functions:
\ba
\VEV{C(z)C(0)} = 0, \quad \VEV{C(z)D(0)} \sim 1, \quad \VEV{D(z)D(0)} \sim -2 \ln (z).
\ea
$O_i$s are the primaries whose four-point functions may contain logarithmic terms, and $C(z), D(z)$ are called logarithmic operators of dimension $h_C$.
In our case, $O_1$ and $O_2$ are $O_{r,t}$ and $O_{q-(r-1),q-(t+2)}$ in (\ref{ex0}), but the {\it r.h.s.}\ of equation should take a non-chiral form. In any case, the underlying theory on the worldsheet is logarithmic CFT as such, and hence minimal string theories turn out to be logarithmic.

Substituting $(r,t)=(2,1)$ into eq.\ (\ref{result0}), one can obtain:
\ba
&\!\!\! &\!\!\! \hspace{-30pt}
I^{(1)} \left( \txt{ \frac{p-2 q}{p},\frac{p-2 q}{p},\frac{2 q-2 p}{p}}; 0; \xi,\bar \xi \right)
\nonumber \\
\hspace{-25pt} &\!\!\! = &\!\!\! 
- \frac{\pi^2}{\sin \left(2\pi \frac{q}{p}\right)}
\nonumber \\
\hspace{-25pt} &\!\!\! {} &\!\!\! 
\times \left\{ \textstyle{ {}_2 F_1 \left( -\frac{p-2q}{p},\frac{2p-2q}{p};1;\xi \right) {}_2 F_1 \left( -\frac{p-2q}{p},\frac{2p-2q}{p};1; 1-\bar \xi \right)} + (c.\,c.)\right\}.
\label{result1}
\ea
In the case of $(p,q)=(4,3)$, one can easily confirm that 
eq.\ (\ref{result1}) reproduces the result in \cite{SY02}. 
Thus, we obtained from the formula (\ref{intrep}) 
logarithmic four-point correlation functions in Liouville field theory 
without applying the differential regularisation procedure \cite{SY02}. 
For more details of the procedure, see also \cite{SY00}.


\newsection{
Conclusions and remarks
}
As the simplest examples of minimal string theory, we have studied the $(p,q)$ minimal models coupled to Liouville field theory. Extracting only the Liouville sector, the four-point functions of tachyons were reduced to two-dimensional integrals of the Coulomb gas type. By using the transformed version of Dotsenko's formula (\ref{intrep}), it was shown that for $p\geq 4$ and $q\geq 3$, certain pairs of tachyons possess logarithmic singularities in their four-point functions, and that minimal string theories are therefore logarithmic. At the end of the previous section, we also confirmed that the previous results \cite{SY02,SY00} can be justified without the regularisation procedure.

It is remarkable that the result in (\ref{result0}) does not require any specific conditions for logarithms, and that the logarithms in (\ref{log}) emerge so naturally. In ordinary free field realisations of CFT, there should be conditions for logarithms which are not necessarily necessary and sufficient.
For example, as shown in \cite{Ishimoto:2003nb}, in the Coulomb gas picture of the minimal models, there are a necessary condition and a necessary and sufficient condition on $(r,t)$ and $(p,q)$ for logarithms in a certain correlation function. The free boson realisation of $SU(2)_k$ WZNW models also possesses restrictive conditions for logarithms.
Unlike such cases,
the result in (\ref{result0}) does not require any restrictive bounds for logarithms, as one can see in the logarithmic expansion (\ref{log}) of the hypergeometric functions. 
This is a remarkable feature of Liouville field theory and $(p\geq 4,\, q\geq 3)$ minimal string theories.

As already implied, our discussion is not restricted to the unitary minimal matters of $(p=q+1,\, q\geq 3)$, but also holds for generic integer values of $(p,q)$, except for few cases.
The only restriction on $(p,q)$ comes from no-pole condition $A\not\in {\mathbf Z}$, which is intrinsically the same as $(1+s)\frac{q}{p} \not\in {\mathbf Z}$ in \cite{Ishimoto:2003nb}. Nonetheless, this is not so restrictive in our case, because the bound, $1\leq t \leq p-3$, with coprime $(p,q)$ automatically satisfies the condition. In addition, if any correlation functions of the type in (\ref{ex0}) exist, the expansion (\ref{log}) tells us that it has logarithms and the theory is said to be logarithmic. Hence, $p,q$ can take quite general integer values, including even non-coprime integers.
 This generality of our derivation may help with understanding the conjecture by Seiberg and Shih that all values of $(p,q)$ correspond to some minimal string theory or deformations thereof \cite{Se04}.

An operator-valued relation for the logarithmic degenerate fields is shown in \cite{Za03}.
The fields appearing in his relation are nothing but puncture-type operator $\phi V_\alpha$, or equivalently $\frac12 \frac{\partial}{\partial \alpha} V_\alpha$.
It is therefore straightforward, as demonstrated by Kogan and Lewis \cite{Kogan:1997fd} that the correlation functions involving such fields yield logarithmic singularities.
Our primaries in the Liouville sector are equivalent to the primaries appearing on the {\it r.h.s.}\ of the relation and therefore can be regarded as singular vectors of the logarithmic degenerate fields, $\phi V_\alpha.$\footnote{\ 
See for example \cite{Gaberdiel:1996kx} for the singular vectors of $c_{p,1}$ models.
}\ \,It should be stressed here that our calculation shows that the correlation functions of such primaries can be explicitly calculated, and that they also yield logarithmic singularity in the correlation functions.
Furthermore, it may help to calculate the correlation function involving the logarithmic degenerate fields by using an inverse relation of Zamolodchikov's operator relation. This would be an interesting application.

In this paper, we have not discussed the effects of boundary, that is, boundary conditions, boundary states, etc. As has already been mentioned briefly, Seiberg and Shih recently discussed the relationship of minimal string theory with matrix models and their branes \cite{Se04}. This has not been dealt with here. These are all interesting directions for further study.

\vspace{30pt}
\noindent
{\large\bf Acknowledgements}
\vspace{10pt}

The authors thank Y.~Tanii and K.~Ohta for their useful discussions. Y.~I. also thanks N.~Kikuchi for his help, and
S.~Y. thanks A.~Nakamula for his comments.




\end{document}